\documentclass[  
 aps,
]{revtex4}
\usepackage{graphicx}
\usepackage[colorlinks=true, pdfstartview=FitV, linkcolor=blue, citecolor=red, urlcolor=magenta]{hyperref}
\usepackage{graphicx}
\usepackage{latexsym}
\usepackage{amsmath}
\usepackage{amsfonts}
\usepackage{amssymb}
\usepackage{verbatim}
\usepackage{braket}
\usepackage{extarrows}
\usepackage{hyperref}

\begin{document}

\title{Vacuum energy density from a self-interacting scalar field in a Lorentz-violating
Ho{\v{r}}ava-Lifshitz model}
\author{$^{1}$A. J. D. Farias Junior}
\email{antonio.farias@ifal.edu.br}
\author{$^{2}$E. R. Bezerra de Mello}
\email{emello@fisica.ufpb.br}
\author{$^{2}$Herondy Mota}
\email{hmota@fisica.ufpb.br}
\affiliation{$^{1}$Instituto Federal de Alagoas,\\
CEP: 57460-000, Piranhas, Alagoas, Brazil} 
\affiliation{$^{2}$Departamento de F\'{\i}sica, Universidade Federal da Para\'{\i}ba,\\
Caixa Postal 5008, Jo\~{a}o Pessoa, Para\'{\i}ba, Brazil}

\begin{abstract}
In this paper we consider a massive self-interacting scalar quantum field in
a Lorentz-violation scenario based on a Ho{\v{r}}ava-Lifshitz model.
Specifically, we investigate the vacuum energy density and its loop correction, up
to first order in the self-interaction coupling constant, and also the
topological mass generation. These quantities are also analyzed in the case
where the field is massless. The scalar vacuum state is perturbed by the
presence of two large parallel plates, placed at a distance $L$ from each
other, due to the imposition of Dirichlet boundary condition on the two
plates. To perform this study, the effective potential approach in quantum
field theory is applied.

\end{abstract}

\maketitle

\section{Introduction}

 One of the most important experimentally confirmed  prediction in quantum field theory, is the Casimir effect. This phenomenon was first established in 1948 \cite{casimir1948attraction}, by considering electromagnetic quantum field confined in a region between two large parallel plates. Although being electrically neutral, theoretically, these plates are attracted to each other as consequence of modification of the quantum vacuum fluctuations due to the boundary condition imposed on the field. This prediction was  experimentally analyzed in \cite{Sparnaay:1958wg}, followed by high precisions experiments \cite{bressi2002measurement, kim2008anomalies,
lamoreaux1997demonstration,lamoreaux1998erratum,mohideen1998precision,mostepanenko2000new,wei2010results}. Casimir-like effects can emerge also by considering other kind of quantum field, such as, scalar  and fermionic fields, obeying specific boundary conditions \cite{romeo2002casimir,aleixo2021thermal,Escobar:2023hzz} (For an extensive review of Casimir effect see \cite{bordag2009advances, milton2001casimir,mostepanenko1997casimir}).

Since the Theory of Relativity is the basis for quantum field theory, the
standard approach for the investigation of the Casimir effect is assuming
that the Lorentz symmetry is preserved. However, high energy scale theories
fail to preserve the Lorentz symmetry \cite{Kos,PhysRevD.55.6760}. In a
scenario where the Lorentz symmetry violation is allowed, the spacetime
becomes anisotropic, modifying  the modes of the quantum field
and as a consequence, the vacuum energy density is affected. The Lorentz symmetry
violation is an interesting topic which has been attracting a great deal of attention, mainly
because it is an alternative to investigate new physics beyond the
standard model. In Ref.~\cite{Aj} the authors considered a scalar field
under helix boundary condition in a scenario with a CPT-even aether-type Lorentz symmetry violation
to study vacuum energy density. Thermal corrections to the vacuum energy density in a
Lorentz-breaking scalar field theory is considered in Ref.~\cite{article}.
Considering finite temperature and an external magnetic field, the
corrections to the vacuum energy due to the Lorentz violation is
investigated in \cite{Santos:2022tbq}. In Ref.~\cite{dantas2023bosonic} it
was considered the Casimir like-effect with Lorentz symmetry violation in a
theory with high order space derivatives.  Moreover, the Lorentz symmetry violation was also considered in string theory \cite{Kos} context, and in low-energy scale
in Refs.~\cite{PhysRevD.55.6760,PhysRevD.58.116002,Ani,Carl,Hew,Ber,Kost,Anc,Ber2,Alf,Alf2}.

The unification of General Relativity with Quantum Mechanics remains one of the major challenges in contemporary physics. In \cite{hovrava2009quantum}, P. Horava proposed a formalism, named Horava-Lifshitz (HL) model, as an attempt to construct a renormalizable quantum field theory for gravity. In this formalism, the propagator for  gravitons  depends on the energy scale, introducing an anisotropy between space and time coordinates. In addition, in \cite{anselmi2009weighted} and \cite{anselmi2009weighted2}, D. Anselmi also proposed a model  that violates Lorentz symmetry explicitly at high energies and is renormalizable by weighted power counting. The model contains higher space derivatives, which improve
the behavior of propagators at large momenta, but no higher time derivatives.  Although being different approaches, both formalism present the same feature, an explicit Lorentz violation symmetry in the higher energy scale. In this paper, even without introducing a direct coupling between the scalar field and gravity,  we will refer to this formalism as being HL-like one.

The analysis of the Casimir energy associated with massless scalar quantum field confined in the region between two large and parallel plates, have been developed in \cite{Ferrari:2010dj} and \cite{ulion2015casimir}, in the context of HL Lorentz violation. More recently, considering massive quantum field, this analysis has been developed in \cite{maluf2020casimir}. In this present analyzes we decided to revisit this previous investigation, i.e., the study of the vacuum energy density associated with the massive scalar quantum field. Our main objectives are to generalize results obtained in \cite{maluf2020casimir}, considering in this model additional $\lambda\varphi^{4}$ self-interaction of the scalar field, and not least important, to clarify some results find there.  In this model we will impose that the field obeys  Dirichlet boundary condition on two large and parallel plates separated by distance $L$. Using the path integral approach to obtain the effective potential \cite{jackiw1974functional}, we develop the investigation of the vacuum energy density and its loop correction in the Horava-Lifshitz formalism. Furthermore, the topological mass which arises due to the boundary condition and also due to the Lorentz symmetry violation is investigated.  Although the breaking of Lorentz symmetry takes place only at Planck energy scales, residual signatures of spacetime anisotropy may be observable in phenomena at lower energy scales, such as in the study of vacuum energy effects.

This paper is organized as follows: in Sec.\ref{sec2} we review the main
aspects of the Ho{\v{r}}ava-Lifshitz model, including a self-interaction
potential, and the path integral formalism that we apply for the
investigation we develop here, as well as the generalized zeta function
technique to write the path integral in a convenient form. In Sec.\ref{sec3}%
, we impose Dirchlet boundary condition on the quantum field and obtain the
generalized zeta function, which allows us to write the
one-loop correction and the effective potential. Sec.\ref{seccas} is
dedicated to the analysis of the vacuum energy density for the case of both
massive and massless fields, which shows a dependence on the boundary
conditions obeyed by the fields and also on the Lorentz symmetry violation.
In addition, the correction for the vacuum energy density is also
considered in this section. Next, in Sec.\ref{sectop}, we investigate the possibility of a
topological mass arising due to the boundary condition, self-interaction
potential and the Lorentz violation within the Ho{\v{r}}ava-Lifshitz
formalism. In Sec.\ref%
{seccomp} we estimate the ratio between the parameter associated with the
Lorentz violation and the length separation $L$ between the plates. Finally in Sec.\ref{sec5} we present our conclusions. Throughout this paper we use natural
units in which both the Planck constant and the speed of light are set equal
to unity, $\hslash =c=1$.

\section{Ho{\v{r}}ava-Lifshitz theory and the one-loop correction}

\label{sec2}

We first consider a massive real scalar quantum field, in a theory where the
space and time have different properties by rescaling. This difference
provides a spacetime anisotropy and as a consequence the Lorentz symmetry
violation. The spacetime anisotropy, makes the theory
invariant under the following scale transformation:
\begin{equation}
x\rightarrow bx,\ \qquad\qquad\qquad t\rightarrow b^{\xi }t,
\end{equation}%
where $\xi $ is a critical exponent \cite{hovrava2009quantum}. In fact the Lorentz invariance is broken for $\xi\neq 1$; however the main objective for this violation is that for a convenient choice of $\xi$ bigger than unity, the theory becomes free of ultraviolet divergence. Specifically focusing on the theory of gravitation, Horava,  in \cite{hovrava2009quantum} adopted $\xi=3$. To avoid problem related with fractional derivative in this work we will assume $\xi$ being an integer number; nonetheless, as we will see later our results are analytical function of $\xi$, and no restrictions need to be imposed on it.

In the 4-dimensional Euclidean spacetime and considering a self interacting
field $\varphi$, we can write the action describing the system as follows \cite%
{anselmi2009weighted, anselmi2009weighted2},%
\begin{equation}
S\left( \varphi \right) =-\frac{1}{2}\int d^{4}x\ \varphi \left[ -\partial
_{\tau }^{2}+\left( -1\right) ^{\xi }l^{2\left( \xi -1\right) }\left(
\partial _{x}^{2}+\partial _{y}^{2}+\partial _{z}^{2}\right) ^{\xi }\right]
\varphi -\int d^{4}x\ V\left( \varphi \right) \,,  \label{s}
\end{equation}%
where%
\begin{equation}
V\left( \varphi \right) =\frac{1}{2}m^{2}\varphi ^{2}+\frac{\lambda }{4!}%
\varphi ^{4}\,,
\end{equation}%
includes the mass and the self-interaction of the field. The parameter $l$
in the above expression has dimension of length, and has been introduced to
make the dimension of the Lagrangian density compatible. Note also that $m$ is the mass of
the field and $\lambda $ is the coupling constant of the self-interaction. Our main motivation in considering the above system, is to generalize previous results developed in \cite{maluf2020casimir}, where the authors consider massive scalar field in a HL-like model. In this present analysis we will show that the vacuum energy density vanishes for the case where $\xi$ is an even number, besides we have also included a $\lambda\phi^4$ self-interaction term. This term introduces correction to the vacuum energy density, and is responsible for the generation of topological mass. Note that we have neglected gravity effects. In fact, our analysis presents itself as a toy model to get a grasp of how a more
realistic interacting model, e.g., the electromagnetic theory with the inclusion of Lorentz violation, is supposed to be constrained. 

Furthermore, in Sec.\ref{sec7}, we make use of experimental data from Ref. \cite{bressi2002measurement} to constrain the 
critical exponent. The analysis of Ref. \cite{bressi2002measurement} does not make any reference to the nature of the quantum field, although the data better fits the electromagnetic field vacuum modes producing the Casimir effect. In spite of that, the bar error associated with the analysis makes possible to include other effects such as the contribution of other quantum fields, finite temperature effects, plates roughness, and so on. This would hep to improve the obtained constraints in Ref. \cite{bressi2002measurement}.

In order to analyze the vacuum energy and the generation of topological mass associated with the system under consideration, we will construct the effective potential by expanding the action about the background field, $\Phi$, by setting $\varphi =\Phi +\phi $, with $\phi $ representing the quantum fluctuations. This provides a series expansion as shown bellow,
\begin{eqnarray}
	\label{Expansion}
	V_{\mathrm{eff}}\left( \Phi \right) =\sum_{n}V^{(n)}(\Phi) \  .
\end{eqnarray}
In our analysis we are interested up to the second order in the expansion. Note that, in this paper, we are adopting natural units.

As it is well known the zero order term in \eqref{Expansion} corresponds to the tree level contribution, and the first and second ones to the 1- and 2-loop corrections, respectively. For a more detailed explanation about this formalism, see
\cite{jackiw1974functional, ryder1996quantum,toms1980symmetry,cruz2020casimir,porfirio2021ground,PhysRevD.107.125019}. 
 The first order contribution, $V^{\left( 1\right) }\left( \Phi \right)$,  can be written in terms of path
integral as:
\begin{equation}
V^{\left( 1\right) }\left( \Phi \right) =-\frac{1}{\Omega _{4}}\ln \int 
\mathcal{D}\varphi \ \exp \left[ -\frac{1}{2}\int d^{4}x\ \varphi \left(
x\right) \hat{A}\varphi \left( x\right) \right] ,
\end{equation}%
where $\Omega _{4}$ is the 4-dimensional volume of the Euclidean spacetime,
which depends on the conditions imposed on the fields.  In our case the operator $\hat{A}$ is given by,
\begin{equation}
\hat{A}=-\partial _{\tau }^{2}+\left( -1\right) ^{\xi }l^{2\left( \xi
-1\right) }\left( \partial _{x}^{2}+\partial _{y}^{2}+\partial
_{z}^{2}\right) ^{\xi }+V^{\prime \prime }\left( \Phi \right) .  \label{a}
\end{equation}%
The double prime notation in $V^{\prime \prime }\left( \Phi \right) $,
stands for the second derivative of $V\left( \varphi \right) $ with respect
to the field $\varphi $ calculated at $\Phi $, which is the fixed background
field. Then, for the theory we consider, $V^{\prime \prime }\left( \Phi
\right) $ takes the following form%
\begin{equation}
V^{\prime \prime }\left( \Phi \right) =m^{2}+\frac{\lambda }{2}\Phi ^{2}.
\label{v}
\end{equation}

 Regarding the one loop correction to the effective potential we will use the zeta-function technique to perform the calculations. Thus, denoting by $\Lambda_\sigma$ the set of the eigenvalues of the operator $\hat{A}$, the corresponding zeta function reads:
\begin{equation}
	\zeta \left( s\right) =\sum_{\sigma }\Lambda _{\sigma }^{-s},  \label{zg}
\end{equation}%
In the above expression $\sigma$ represents the complete set of quantum numbers that characterize the eigenfunction of $\hat{A}$. The summation symbol denotes sum over discrete quantum numbers, and integration over continuous ones. Once we construct the generalized zeta function in Eq.~(\ref{zg}), we write the one-loop correction to the effective potential in
the following form \cite{toms1980symmetry, PhysRevD.107.125019,
hawking1977zeta}:
\begin{equation}
V^{\left( 1\right) }\left( \Phi \right) =-\frac{1}{2\Omega _{4}}\left[ \zeta
^{\prime }\left( 0\right) +\zeta \left( 0\right) \ln \nu ^{2}\right] .
\label{1l}
\end{equation}%
The notations $\zeta \left( 0\right) $ and $\zeta ^{\prime }\left( 0\right) $
stand for the generalized zeta function and its derivative with respect to $%
s $, evaluated at $s=0$, respectively. The parameter $\nu $ has dimension of
mass which is to be removed by a renormalization procedure.

At this point we would like to mention that, although we have used an Euclidean extension of the Klein-Gordon operator, Eq. \eqref{a}, the effective potential obtained is a real function. Thus, the mathematical approach adopted here, does not affect the physical result.

 On the other hand, in order to calculate the two-loop correction to the effective potential we will use Feynman graphs. As we are interested only in the vacuum contribution, this correction can be written in the terms of the generalized zeta function \cite%
{cruz2020casimir,porfirio2021ground,PhysRevD.107.125019}. We postpone the
explicit form of $V^{\left( 2\right) }\left( \Phi \right) $ until Section \ref{sec4C}.
\section{Dirichlet boundary condition, generalized zeta function and the
one-loop correction}
\label{sec3}

In this section, we apply the Dirichlet boundary condition on
two parallel large plates, separated by a distance $L$ along the $z$ direction.
Within this configuration, the scalar field $\varphi $ satisfies the
following restriction:%
\begin{equation}
\varphi \left( \tau ,x,y,0\right) =\varphi \left( \tau ,x,y,L\right) =0.
\end{equation}%
The above condition implies that the eigenvalues of the operator $\hat{A}$
given in Eq.~(\ref{a}) takes the form,%
\begin{equation}
\Lambda _{\sigma }=k_{\tau }^{2}+\left( -1\right) ^{2\xi }l^{2\left( \xi
-1\right) }\left[ k_{x}^{2}+k_{y}^{2}+\left( \frac{\pi n}{L}\right) ^{2}%
\right] ^{\xi }+V^{\prime \prime }\left( \Phi \right) .  \label{e}
\end{equation}%
Note that the momentum in $z$ has been discretized and $%
n $ assumes non-negative integers values, that is, $\ n=1,2,3,...$\;. The set of quantum modes 
in this case is given by $\sigma=(k_{\tau}, k_x, k_y, n)$, with ($k_{\tau}, k_x, k_y$) being the continuous momenta.

The eigenvalues given in Eq.~(\ref{e}) are used to construct the generalized
zeta function (\ref{zg}), which is written as,%
\begin{equation}
\zeta \left( s\right) =\frac{\Omega _{3}}{\left( 2\pi \right) ^{3}}%
\sum_{n=1}^{\infty }\int dk_{\tau }dk_{x}dk_{y}\left\{ k_{\tau }^{2}+\left(
-1\right) ^{2\xi }l^{2\left( \xi -1\right) }\left[ k_{x}^{2}+k_{y}^{2}+%
\left( \frac{\pi n}{L}\right) ^{2}\right] ^{\xi }+V^{\prime \prime }\left(
\Phi \right) \right\} ^{-s}.
\label{momentaI}
\end{equation}%
The quantity $\Omega _{3}$ represents the 3-dimensional volume associated
with the Euclidean time coordinate $\tau $ and the spatial coordinates $x$
and $y$. In order to obtain the generalized zeta function in a more
convenient form, we start by using the following identity:

\begin{equation}
w^{-s}=\frac{2}{\Gamma \left( s\right) }\int_{0}^{\infty }dt\
t^{2s-1}e^{-wt^{2}},
\end{equation}%
where $\Gamma(s)$ is the gamma function. This allows us to perform the Gaussian-like integral in the $k_{\tau }$ variable present in Eq. \eqref{momentaI}, providing that,%
\begin{equation}
\zeta \left( s\right) =\frac{2\Omega _{3}\pi ^{\frac{1}{2}}}{\left( 2\pi
\right) ^{3}\Gamma \left( s\right) }\sum_{n=1}^{\infty }\int
dk_{x}dk_{y}\int_{0}^{\infty }dt\ t^{2s-2}\exp \left\{ -bt^{2}\left[ \left(
k_{x}^{2}+k_{y}^{2}+\left( \frac{\pi n}{L}\right) ^{2}\right) ^{\xi }+\frac{%
V^{\prime \prime }\left( \Phi \right) }{b}\right] \right\} ,
\end{equation}%
where we have defined the quantity $b$ as,%
\begin{equation}
b=l^{2\left( \xi -1\right) }.  \label{b}
\end{equation}%
At this stage, it is more appropriate to write the $k_{x}$ and $k_{y}$ integrals
in polar coordinates and perform the integral in the polar angle. After we
perform two changes of integration variables, namely, $\chi=\frac{b\pi^{2\xi}}{L^{2\xi}}t^{2}$
and $\omega =\chi\left[ \left( \frac{kL}{\pi }\right) ^{2}+n^{2}\right] ^{\xi }$,
we end up with the following expression for the generalized zeta function:%
\begin{equation}
\zeta \left( s\right) =\frac{\Omega _{3}b^{\frac{1}{2}-s}\pi ^{\frac{1}{2}%
-2\xi s+\xi }}{8\xi L^{2-2\xi s+\xi }\Gamma \left( s\right) }%
\sum_{n=1}^{\infty }\int_{0}^{\infty }d\chi\ \chi^{s-\frac{3}{2}-\frac{1}{\xi }%
}e^{-\chi U}\int_{\chi n^{2\xi }}^{\infty }d\omega \ \omega ^{\frac{1}{\xi }%
-1}e^{-\omega },  \label{gz1}
\end{equation}%
where we set $U$ in the form, 
\begin{equation}
U=\frac{L^{2\xi }}{b\pi ^{2\xi }}V^{\prime \prime }\left( \Phi \right) .
\label{u}
\end{equation}

In the generalized zeta function given in Eq.~(\ref{gz1}), we identify the
incomplete gamma function $\Gamma \left( \alpha ,x\right) $ defined as \cite%
{gradshteyn2014table}, 
\begin{equation}
\Gamma \left( \alpha ,x\right) =\int_{x}^{\infty }dt\ t^{\alpha -1}\
e^{-t}=e^{-x}\Psi \left( 1-\alpha ,1-\alpha ,x\right) ,\ \ \ \ \ \ \ \ \text{%
Re\ }\alpha >0,
\end{equation}%
where the function $\Psi \left( \alpha ,\beta ,z\right) $ is a combination
of confluent hypergeometric functions $_{1}F_{1}\left( \alpha ,\gamma
,z\right) $ \cite{gradshteyn2014table}, that is,%
\begin{equation}
\Psi \left( \alpha ,\beta ,z\right) =\frac{\Gamma \left( 1-\gamma \right) }{%
\Gamma \left( \alpha -\gamma +1\right) }\,_{1}F_{1}\left( \alpha ,\gamma
,z\right) +\frac{\Gamma \left( 1-\gamma \right) }{\Gamma \left( \alpha
\right) }z^{1-\gamma }\,_{1}F_{1}\left( \alpha -\gamma +1,2-\gamma ,z\right)
.
\end{equation}%
Hence, we obtain the following expression for the generalized zeta function:%
\begin{equation}
\zeta \left( s\right) =\frac{\Omega _{3}b^{\frac{1}{2}-s}\pi ^{\frac{1}{2}%
-2\xi s+\xi }}{8\xi \Gamma \left( s\right) L^{2-2\xi s+\xi }}%
\sum_{n=1}^{\infty }\int_{0}^{\infty }d\tau \ \frac{\tau ^{s-\frac{3}{2}-%
\frac{1}{\xi }}}{n^{2\xi s-\xi -2}}\ e^{-\tau \left( 1+\frac{U}{n^{2\xi }}%
\right) }\ \Psi \left( 1-\frac{1}{\xi },1-\frac{1}{\xi },\tau \right),
\label{gz2}
\end{equation}%
where we have made another change of variable, namely, $\tau =tn^{2\xi }$%
. The form of the generalized zeta function above suggests the use of the
integral below \cite{gradshteyn2014table}, i.e.,
\begin{equation}
\int_{0}^{\infty }t^{b-1}\ e^{-zt}\ \Psi \left( a,c,t\right) dt=\frac{\Gamma
\left( b\right) \Gamma \left( b-c+1\right) z^{-b}}{\Gamma \left(
a+b-c+1\right) }\,_{2}F_{1}\left( a,b,a+b-c+1,1-z^{-1}\right) \ .  \label{g}
\end{equation}%
Therefore, with the help of Eq.~(\ref{g}), we can rewrite the generalized
zeta function, Eq.~(\ref{gz2}), as 
\begin{eqnarray}
&&\zeta \left( s\right) =\frac{\Omega _{3}\pi ^{\frac{1}{2}-2\xi s+\xi }}{%
8\xi L^{2-2\xi s+\xi }}\frac{b^{\frac{1}{2}-s}\Gamma \left( s-\frac{1}{2}-%
\frac{1}{\xi }\right) \Gamma \left( s-\frac{1}{2}\right) }{\Gamma \left(
s\right) \Gamma \left( s+\frac{1}{2}-\frac{1}{\xi }\right) }  \notag \\
&&\times \sum_{n=1}^{\infty }\left( n^{2\xi }+U\right) ^{\frac{1}{2}+\frac{1%
}{\xi }-s}\,_{2}F_{1}\left( 1-\frac{1}{\xi },s-\frac{1}{2}-\frac{1}{\xi },s+%
\frac{1}{2}-\frac{1}{\xi },\frac{U}{n^{2\xi }+U}\right) ,  \label{zh1}
\end{eqnarray}%
where the function $_{2}F_{1}\left( \alpha ,\beta ,\gamma ,z\right) $ is the
hypergeometric function \cite{gradshteyn2014table}. The sum above is clearly divergent so that we need a renormalization method 
to subtract the infinite contribution and only work with the finite part of the generalized zeta function.

In order to perform the sum over the quantum number $n$ in Eq.~(\ref%
{zh1}) and obtain a finite contribution, we use the Abel-Plana summation formula \cite%
{bordag2009advances,PhysRevD.21.933,
ulion2015casimir,saharian2007generalized}. The calculations using this formula are described in Appendix \ref{app1} where it is shown
that there are three resulting terms, but only one provides a
finite vacuum energy density, which is given by%
\begin{eqnarray}
\zeta _{0}\left( s\right) =&-&\frac{\Omega _{3}\pi ^{\frac{1}{2}+\xi -2\xi s}b^{%
\frac{1}{2}-s}}{4\xi L^{2+\xi -2\xi s}}\frac{\Gamma \left( s-\frac{1}{2}-%
\frac{1}{\xi }\right) \Gamma \left( s-\frac{1}{2}\right) \sin \left[ \pi
\left( \frac{\xi }{2}+1-\xi s\right) \right] }{\Gamma
\left( s\right) \Gamma \left( s+\frac{1}{2}-\frac{1}{\xi }\right) }  \notag
\\
&&\times \int_{U^{\frac{1}{2\xi }}}^{\infty }dx\mathcal{F}_{s}\left( \xi ,U,x\right) \ ,  \label{abel3}
\end{eqnarray}
where we have defined the function $\mathcal{F}_{s}\left( \xi ,U,x\right)$ as
\begin{equation}
\mathcal{F}_{s}\left( \xi ,U,x\right) =\frac{\left( x^{2\xi
}-U\right) ^{\frac{1}{2}+\frac{1}{\xi }-s}}{e^{2\pi x}-1}\,_{2}F_{1}\left( 1-%
\frac{1}{\xi },s-\frac{1}{2}-\frac{1}{\xi },s+\frac{1}{2}-\frac{1}{\xi },%
\frac{U}{U-x^{2\xi }}\right) .
\label{f}
\end{equation}%

With the finite generalized zeta function obtained above, we proceed to calculate the
one-loop correction to the effective potential from Eq.~(\ref{1l}). In order to do so, we have to evaluate the generalized zeta function above and
its derivative at $s=0$, that is, $\zeta \left( 0\right) $
and $\zeta ^{\prime }\left( 0\right) $. From the calculations presented in Appendix \ref{app1},  we find that,%
\begin{eqnarray}
&&\zeta _{0}\left( 0\right) =0,  \notag \\
&&\zeta _{0}^{\prime }\left( 0\right) =
\frac{\Omega _{3}\pi ^{1+\xi }b^{\frac{1}{2}}\sin(\pi\xi/2)}{\left( \xi +2\right) L^{2+\xi
}}\int_{U^{\frac{1}{2\xi }}}^{\infty }dx\ \mathcal{F}_0%
\left( \xi ,U,x\right) ,  \label{zcas}
\end{eqnarray}%
where we should remember that the quantity $U$ is given in Eq.~(\ref{u}) together
with (\ref{v}). We should also emphasize that the contribution presented in Eq.~(\ref{zcas}) is the one
that gives rise to the vacuum energy density, as we shall see.

Due to the renormalization procedure performed using Abel-Plana summation formula \cite%
{ulion2015casimir}, the one-loop correction to the effective potential takes
into consideration only the contribution given in Eq.~(\ref{zcas}),
therefore the one-loop correction, Eq.~(\ref{1l}), is written in the
following manner:%
\begin{equation}
V^{\left( 1\right) }\left( \Phi \right) =-\frac{\pi ^{1+\xi }l^{\xi -1}\sin(\pi\xi/2)}{2\left( \xi +2\right) L^{3+\xi }}%
\int_{U^{\frac{1}{2\xi }}}^{\infty }dx\ \mathcal{F}_0\left( \xi ,U,x\right)
\,.  \label{vcas}
\end{equation}%
Hence, the renormalized effective potential for a massive self-interacting
scalar field within the Ho{\v{r}}ava-Lifshitz formalism takes the form
\begin{eqnarray}
V_{\text{\textrm{eff}}}^{\mathrm{R}}\left( \Phi \right) =\frac{1}{2}m\Phi
^{2}+\frac{\lambda }{4!}\Phi ^{4}-\frac{\pi ^{1+\xi }l^{\xi -1}\sin(\pi\xi/2)}{2\left( \xi +2\right) L^{3+\xi }}\int_{U^{\frac{1}{2\xi }%
}}^{\infty }dx\ \mathcal{F}_0\left( \xi ,U,x\right) \,.  \label{vr}
\end{eqnarray}
 An interesting aspect to be noted in our result is that it has not been necessary to introduce counterterms in the action \eqref{s} to obtain the renormalized effective potencial above, in contrast with Ref. \cite{cruz2020casimir} in the case $\xi=1$. In the latter, besides $\zeta _{0}(0)\ne 0$, $\zeta _{0}^{\prime }(0)$ contains logarithmic divergence when $\Phi=0$ and also a dependence on the scale $\nu$ present here in Eq. \eqref{1l}. This kind of terms need to be removed which, in the context of the effective potential approach, is usually done by introducing counterterms in the action. The absence of this logarithmic divergence in our case is a characteristic of the H-L model adopted. Thus, the counterterms are needless. Although we have considered $\xi$ being an integer number, our result above is an analytical function of it; consequently by an analytical continuation Eq. \eqref{vr} remains valid for any value of $\xi$.

Once we obtain the explicit form of the renormalized effective potential, up
to one-loop correction, we can calculate the vacuum energy density and also
analyze a possible generation of topological mass. Let us do this in the proceeding sections.

\section{Vacuum energy density and topological mass}

\label{seccas}In this section we investigate the vacuum energy density for a real scalar field, its corresponding order $\lambda$ correction, which
is in fact the contribution from the two-loop correction to the effective potential calculated at
the vacuum state, and also the topological mass. Next, we consider first the vacuum energy density.

\subsection{Vacuum energy density for the massive field}

Upon using the renormalized effective potential \eqref{vr} obtained in the previous section, we
can calculate the vacuum energy density directly by taking the vacuum state,
i.e., $\Phi =0$. Hence, we obtain the vacuum energy density in the following form:%
\begin{eqnarray}
\mathcal{E}_{0}&=&V_{\text{\textrm{eff}}}^{R}\left( 0\right)\nonumber\\
&=&-\frac{\pi\sin(\pi\xi/2)}{2L^4\left( \xi +2\right) R_{\xi}}\int_{U_{0}^{\frac{1}{%
2\xi }}}^{\infty }dx\ \mathcal{F}_0\left( \xi ,U_{0},x\right) ,  \label{cas}
\end{eqnarray}%
where the function $\mathcal{F}_0\left( \xi ,U_{0},x\right) $ is presented
in Eq.~(\ref{f}), with $U_{0}$ given by,%
\begin{eqnarray}
U_{0}=\left. U\right\vert _{\Phi =0}&=& (mL)^2\left[\frac{1}{\pi^{\xi}}\left(\frac{L}{l}\right)^{\xi-1}\right]^2\nonumber\\
&=& (mL)^2R_{\xi}^2.
\label{u01}
\end{eqnarray}%
As we can see, the vacuum energy density given in Eq.~(\ref{cas}) can be
positive or negative depending on the value of the parameter $\xi $ which, as seen in Appendix \ref{app1},
assumes only odd values. The expression (\ref{cas}) is the
general result for the vacuum energy density, for a massive scalar field
within the Ho{\v{r}}ava-Lifshitz formalism, satisfying Dirichlet boundary
conditions on the parallel plates. Unfortunately, it is not possible to solve the integral in
Eq.~(\ref{cas}) for all odd values of $\xi $, except the standard case $\xi=1$ that preserves the Lorentz symmetry.

Let us now closely consider the standard case in
which the critical exponent is set equal to the unity, i.e., $\xi =1$. Since $F\left( 0,\beta ,\gamma ,x\right) =1$, the vacuum energy
density in Eq.~(\ref{cas}) becomes
\begin{equation}
\mathcal{E}_{0}\left( \xi =1\right) =-\frac{m^4}{6\pi^{2}}%
\sum_{n=1}^{\infty }\int_{1}^{\infty }dy\left( y^{2}-1\right) ^{\frac{3}{2}%
}e^{-2n mLy} ,  \label{cx1}
\end{equation}%
where we have performed the change of integration variable, $y=xU_0^{-\frac{1}{2}}$. Note also that the
sum in $n$ of the exponential above provides the expression $(e^{2mLy}-1)^{-1}$. 

The integral in Eq.~(\ref{cx1}) can be performed with the help of the following
identity \cite{gradshteyn2014table}:%
\begin{equation}
\int_{1}^{\infty }dx\left( x^{2}-1\right) ^{\nu -1}e^{-\mu x}=\pi ^{-\frac{1%
}{2}}\left( \frac{2}{\mu }\right) ^{\nu - \frac{1}{2}}\Gamma \left( \nu \right) K_{\nu - \frac{1}{2}}\left(
\mu \right) ,
\label{iden2}
\end{equation}%
where $K_{2}\left( \mu \right) $ is the Macdonald function or the modified
Bessel function of the second kind \cite{abramowitz1965handbook}. Hence, the
vacuum energy density for the case $\xi =1$ takes the form,%
\begin{equation}
\mathcal{E}_{0}\left( \xi =1\right) =-\frac{m^{2}}{8\pi ^{2}L^{2}}%
\sum_{n=1}n^{-2}K_{2}\left( 2nmL\right) .  \label{cx12}
\end{equation}%
The above result is in agreement with the standard result where the Lorentz symmetry is preserved which we can check, for instance, in Refs.~\cite{PhysRevD.96.045019,
bordag2001new, milton2001casimir}.
\begin{figure}[tbp]
\includegraphics[width=.5\textwidth]{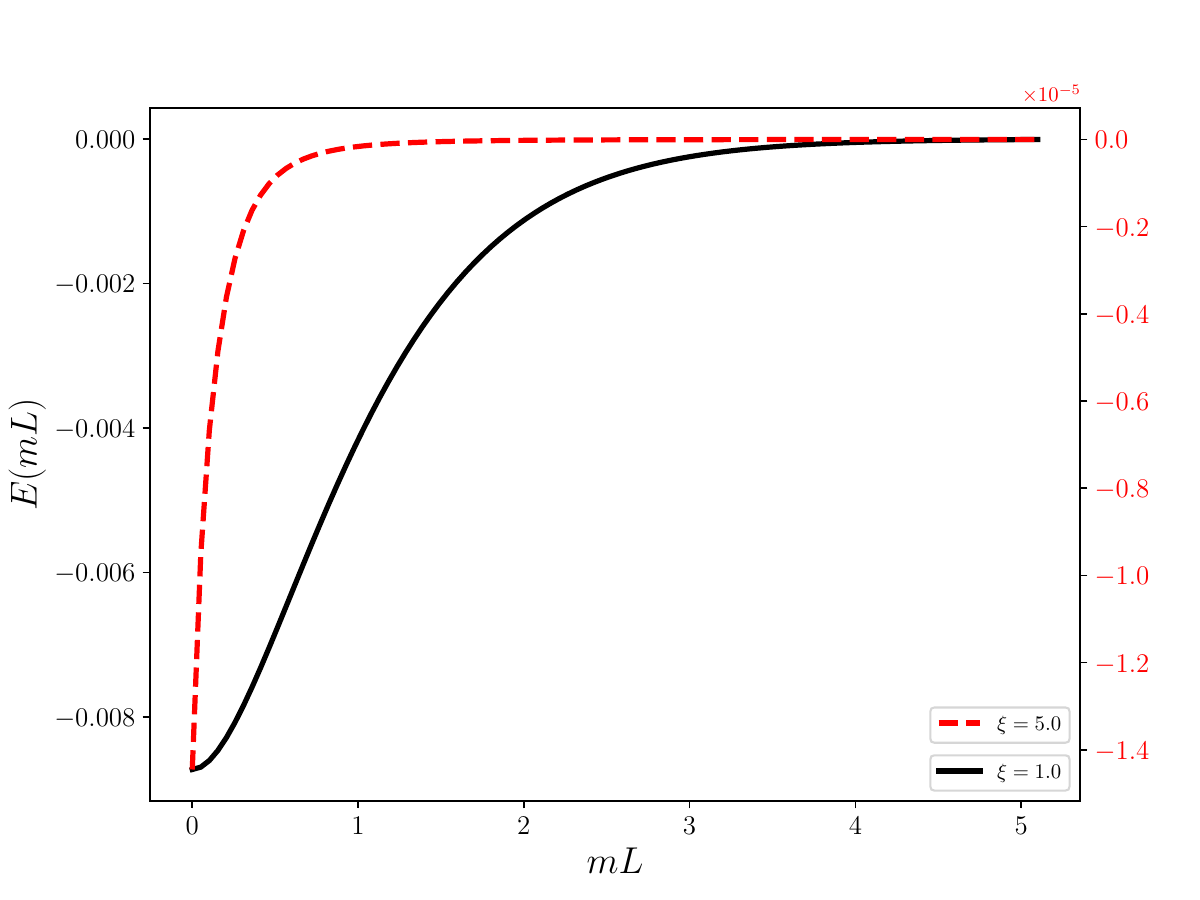}\hfill %
\includegraphics[width=.5\textwidth]{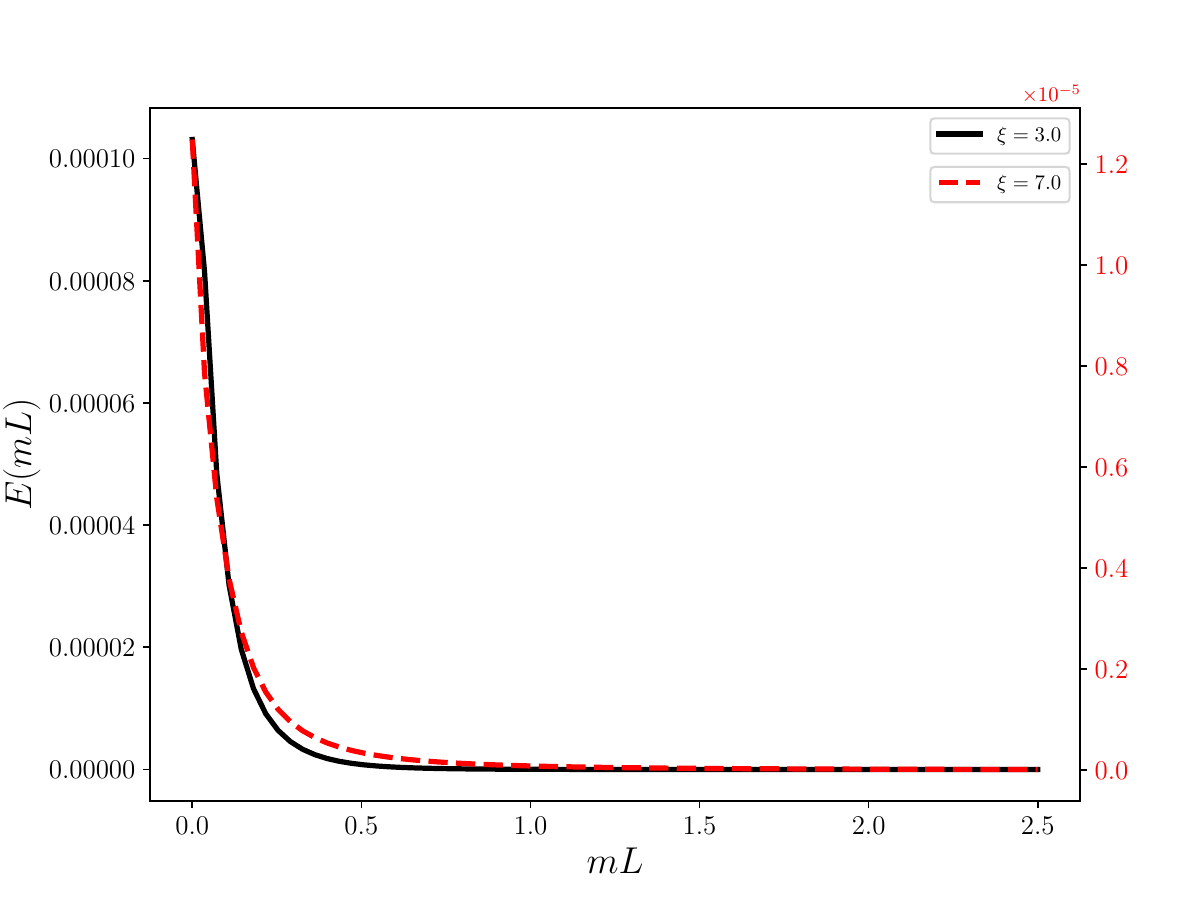}
\caption{Graphs of the dimensionless vacuum energy density $E\left( mL\right)
= \frac{4}{\protect\pi} L^{4}\mathcal{E}_{0}$, as a function of $mL$. The graph on
the left considers the values $\protect\xi =1.0$ (solid black curve) and $%
\protect\xi=5.0$ (dashed red curve), while the plot on the right considers
the values $\xi=3.0$ (solid black curve) and $\xi=7.0$ (dashed red curve). The
scale of each curve is shown in the vertical axes with the correspondent color.  
}
\label{fig1}
\end{figure}

In order to construct a graph for the vacuum energy density given in Eq.~(\ref%
{cas}) as a function of $mL$, it is more convenient to express it in terms of the variable, $z=x^{2\xi }-U$. Hence, Eq. (\ref%
{cas}) can be written as%
\begin{eqnarray}
\mathcal{E}_{0}=-\frac{\pi\sin(\pi\xi/2)}{4L^4\xi \left( \xi +2\right) R_{\xi}}%
\int_{0}^{\infty }dz\frac{z^{\frac{1}{2}+\frac{1}{\xi }}\left( z+U_{0}\right) ^{\frac{1}{2\xi }-1}
}{\left[e^{2\pi (z + U_0)^{\frac{1}{2\xi }}}-1\right]}
\,_{2}F_{1}\left( 1-\frac{1}{\xi },-\frac{1}{2}-\frac{1}{\xi },\frac{1}{2}-%
\frac{1}{\xi },-\frac{U_{0}}{z}\right) \,,
\label{variablez}
\end{eqnarray}%
where $U_0$ has been defined in terms of $R_{\xi}$ in Eq. \eqref{u01}. Thus, we shall now consider graphs of the dimensionless vacuum energy density, $%
E\left( mL\right) =\frac{4L^{4}}{\pi}\mathcal{E}_{0}$, as a function of $mL$. 

In Fig.\ref{fig1} we exhibit graphs of the dimensionless vacuum energy
density $E\left( mL\right) $, as a function of $mL$, considering the values $%
\xi =1$, $3$, $5$, $7$. The graph on the left shows the curves for $\xi =1$ and $%
\xi =5$. The solid black curve stands for the case $\xi =1$ while
the dashed red curve for the case $\xi =5$. Similarly, the graph on the right considers the cases where $\xi =3$
(solid black line) and $\xi =7$ (dashed red line). These plots also show a strong decay of the vacuum energy density with $mL$. Explicitly in \eqref{cx12} we can see an exponential decay for $mL>>1$.  Note that the scale of each curve is correspondently shown on the left and on the right of the vertical axes. Although in section \ref{sec7} we make an estimation on the ratio $\frac{l}{L}$ by using experimental data, for simplicity, we have taken values
of $R_{\xi}$ such that
\begin{equation}
R_{1}=1/\pi ,\ \ \ R_{3}=24,\ \ R_{5}=32,\ \ \ R_{7}=80,\ \ \ R_{9}=260.  \label{r}
\end{equation}%
Therefore, the plot on the left of Fig.\ref{fig1} shows that the order of the vacuum energy density, in absolute values, greatly decreases compared with the 
standard case $\xi=1$. This aspect is also shown in the plot on the right, which also reveals that for values $\xi=3,7$ the vacuum energy density 
gives rise to a repulsive force, in contrast with the cases $\xi=1,5$. 

Let us now consider the massless scalar field case. For this, we should take the limit 
$m\rightarrow 0$ ($U_0\rightarrow 0$) in Eq.~(\ref{cas}). This provides
\begin{equation}
\mathcal{E}_{0}=-\frac{\pi
^{1+\xi }l^{\left( \xi -1\right) }\sin(\pi\xi/2)}{2\left( \xi +2\right) L^{3+\xi }}%
\int_{0}^{\infty }\frac{x^{\xi +2}}{e^{2\pi x}-1}dx.
\end{equation}%
The integral above can be performed by using the
following relation \cite{gradshteyn2014table}:%
\begin{equation}
\int_{0}^{\infty }dx\ \frac{x^{\nu -1}}{e^{\mu x}-1}=\mu ^{-\nu }\Gamma
\left( \nu \right) \zeta _{R}\left( \nu \right) ,\ \ \ \ \ \ \ \ \ \mu >0,\
\ \ \ \ \ \ \ \ \nu >1,  \label{int2}
\end{equation}%
where $ \zeta _{R}(s)$ is the Riemann zeta function. Therefore, we obtain the vacuum energy density for the massless field as%
\begin{equation}
\mathcal{E}_{0}=- \frac{l^{\left(
\xi -1\right) }\sin(\pi\xi/2)}{2^{\xi +4}\pi ^{2}L^{3+\xi }}\Gamma \left( \xi +2\right)
\zeta _{R}\left( \xi +3\right),  \label{cm01}
\end{equation}%
which is in agreement with the result found in Ref. \cite{ulion2015casimir}.

We want now go further to consider the correction to the vacuum energy density, which
is proportional to the self-interaction coupling constant $\lambda $, in linear order. This contribution comes from
the two-loop correction to the effective potencial, as we shall see below.
\subsection{Order-$\lambda$ correction to the vacuum energy density}
\label{sec4C}
The two-loop correction to the effective potential at the vacuum state $\Phi=0$,
provides a order-$\lambda$ loop correction to the vacuum energy density, Eq.~(\ref{cas}). 
This contribution can be written in terms of the finite generalized zeta function \eqref{abel3} as \cite{porfirio2021ground, Aj,
PhysRevD.107.125019},%
\begin{equation}
\mathcal{E}_{0}^{\lambda }=V^{\left( 2\right) }\left( 0\right) =\frac{%
\lambda }{8}\left[ \frac{\zeta _{0}\left( 1\right) }{\Omega _{4}}\right]
_{\Phi =0}^{2},
\end{equation}%
where
\begin{equation}
\left. \frac{\zeta _{0}\left( 1\right) }{\Omega _{4}}\right\vert _{\Phi =0}=-%
\frac{\pi ^{1-\xi }\sin(\pi\xi/2) }{2\left( \xi
-2\right) l^{\left( \xi -1\right) }L^{3-\xi }}\int_{U_{0}^{\frac{1}{2\xi }}}^{\infty }dx\ \mathcal{F}_1\left(\xi
,U_{0},x\right),
\end{equation}%
with $\Omega_4=\Omega_3L$. This leads to the following order-$\lambda$ correction to the vacuum energy density:%
\begin{equation}
\mathcal{E}_{0}^{\lambda }=\frac{\lambda \pi ^{2}R_{\xi}^2}{32L^4\left( \xi -2\right) ^{2}}\left[ \int_{U_{0}^{\frac{1}{2\xi }}}^{\infty }dx\ \mathcal{F}_1\left(\xi
,U_{0},x\right) \right] ^{2},  \label{c1}
\end{equation}%
where $R_{\xi}$ has been defined in Eq. \eqref{u01} and the function $\mathcal{F}_s\left(\xi
,U_{0},x\right) $ in Eq. \eqref{f}. In order to construct a graph for the above correction 
it is useful to use the same change of variable as in Eq. \eqref{variablez}.
\begin{figure}[tbp]
\includegraphics[width=.5\textwidth]{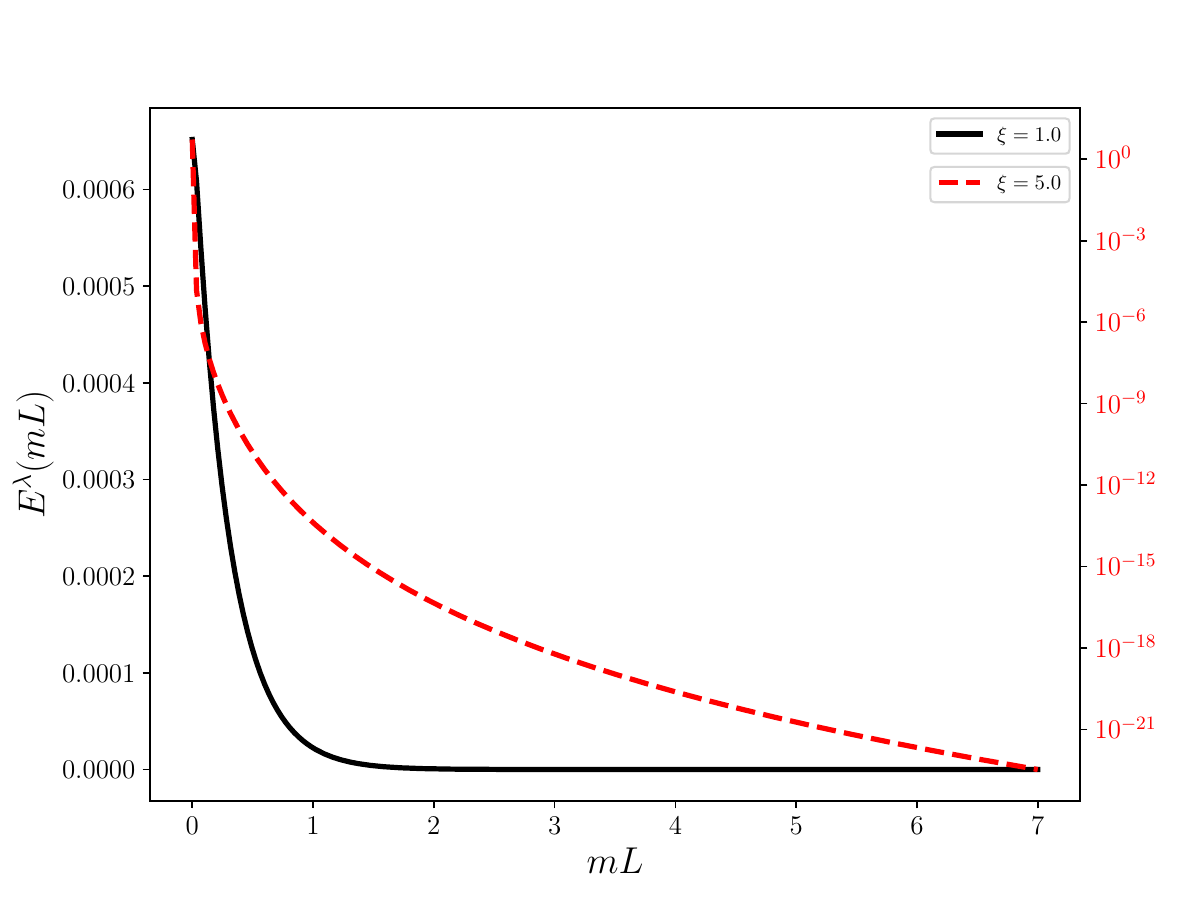}\hfill %
\includegraphics[width=.5\textwidth]{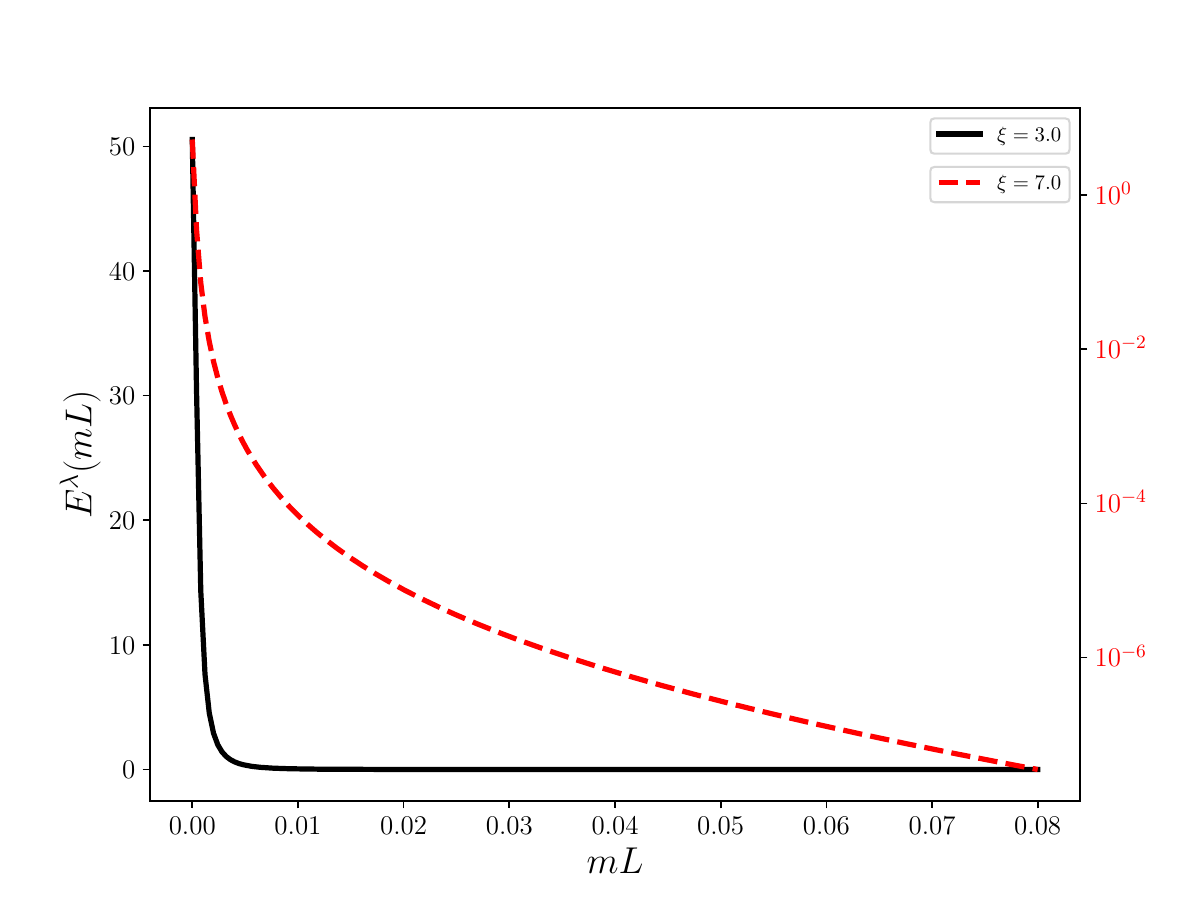}
\caption{Graphs of the dimensionless order-$\lambda$ contribution to the vacuum
energy density $E^{\lambda}\left( mL\right) =\frac{128}{\protect\lambda \protect%
\pi ^{2}}L^{4}\mathcal{E}_{0}^{\protect\lambda }$, as a function of $mL$. The graph on
the left considers the values $\protect\xi =1.0$ (solid black curve) and $%
\protect\xi=5.0$ (dashed red curve), while the plot on the right considers
the values $\xi=3.0$ (solid black curve) and $\xi=7.0$ (dashed red curve). The
scale of each curve is shown in the vertical axes with the correspondent color.}
\label{fig2}
\end{figure}

In Fig.\ref{fig2} we have plotted the dimensionless order-$\lambda$ correction to the
vacuum energy density $E^{\lambda}\left( mL\right) =\frac{128}{\lambda \pi ^{2}}
L^{4} \mathcal{E}_{0} ^{ \lambda } $ as a function of $mL$.
The graph on the left considers the case $\xi=1$ represented by the solid
black curve, while the case $\xi=5$ is represented by the dashed red
curve. Similarly, the graph on the right shows  the cases $\xi=3$ (solid black curve) 
and $\xi=7$ (dashed red curve). For both plots the scale of the vacuum energy density 
is shown in the vertical axes with the correspondent color. Note that in both plots the red vertical axis 
is in logarithmic scale. Note also that the values given in Eq. (\ref{r}) are also
considered here. We can again see that the scale of the vacuum energy density greatly decreases 
in each case, as compared with the standard case $\xi=1$. The graphs above also show that the
correction to the vacuum energy density goes to zero as $mL\rightarrow\infty$. However, exactly at
$mL=0$ (massless case), the correction diverges for $\xi>1$, as we shall see more clearly below.

We turn now to the massless scalar field case, which can be obtained by taking the limit $m\rightarrow 0$ ($U_0\rightarrow 0$)
of Eq. \eqref{c1}. This gives
\begin{eqnarray}
\mathcal{E}_{0}^{\lambda} =\frac{\lambda \pi ^{2-2\xi }}{32\left(
\xi -2\right) ^{2}l^{2\left( \xi -1\right) }L^{6-2\xi }}\left[
\int_{0}^{\infty }dx\ \frac{x^{2-\xi}}{e^{2\pi x}-1}\right] ^{2}.
 \label{c2}
\end{eqnarray}
Note that the integral above does not converge for $\xi>1$. This is a consequence of the Ho{\v{r}}ava-Lifshitz model adopted here, which helps to eliminate ultraviolet divergences but as a consequence brings infrared ones in the massless limit of Eq. \eqref{c1}. The question whether 
this infrared divergence can be treated will be considered elsewhere. 

Fortunately, the integral in Eq. \eqref{c2} converges for $\xi=1$ and provides the standard result where the Lorentz symmetry is preserved \cite{cruz2020casimir}. So, let us first consider $\xi =1$ in Eq. \eqref{c1}. In this case, we obtain the
following result:%
\begin{eqnarray}
\mathcal{E}_{0}^{\lambda }\left( \xi =1\right) &=&\frac{\lambda m^4}{32\pi^4}\left[\sum_{n=1}^{\infty }\int_{1}^{\infty }dy\left( y^{2}-1\right) ^{\frac{1}{2}%
}e^{-2n mLy}  \right]^{2}\nonumber\\
&=&\frac{\lambda m^{2}}{128\pi
^{4}L^{2}}\left[ \sum_{n=1}^{\infty }n^{-1}K_{1}\left( 2nmL\right) \right]
^{2},
\end{eqnarray}%
which is in agreement with the results found in Ref.~\cite{cruz2020casimir} in the case with no
Lorentz violation. Note that we have used again the change of variable $y=xU_0^{-\frac{1}{2}}$, and also Eq. \eqref{iden2}. This is similar to what we have done
in Eq. \eqref{cx1}.

In addition, from Eq. \eqref{c2}, for a massless scalar field, we have%
\begin{equation}
\mathcal{E}_{0}^{\lambda}\left( \xi =1\right) =\frac{%
\lambda }{18432L^{4}},
\end{equation}
which is also in agreement with Ref.~\cite{cruz2020casimir}. 

 In general the zeta-function, $\zeta_0(s)$, evaluated at $s=1$ presents a divergent contribution, as it was pointed out in \cite{cruz2020casimir}. This contribution needs to be  subtracted by some regularization procedure. However, in the H-L context, there is no such divergence, and all results are finite, as we have shown above.

In the next section we consider an analysis for the topological
mass generated by the boundary condition. We, thus, expect the result will
depend on the self-interaction coupling $\lambda$ and the Lorentz violation parameter $l$.

\subsection{Topological mass for the massive field}

\label{sectop}In this section we investigate the topological mass which
arises due to the boundary condition, self-interaction potential and the
Lorentz violation within the Ho{\v{r}}ava-Lifshitz formalism used here. 

The topological mass squared of the field can be written as the second
derivative of the effective potential, evaluate at the vacuum state, that
is, $\Phi =0$ \cite{toms1980symmetry}. Hence, using the effective potential
given in Eq.~(\ref{vr}) we write the topological mass as,%
\begin{equation}
m_{\text{\textrm{T}}}^{2}=\left. \frac{d^{2}V_{\text{\textrm{eff}}}^{\mathrm{%
R}}\left( \Phi \right) }{d\Phi ^{2}}\right\vert _{\Phi =0}\,.  \label{tm01}
\end{equation}%
In order to perform the derivative of the effective potential, we have to
apply the Leibniz rule \cite{arfken2011mathematical}. The final result
reads,%
\begin{eqnarray}
m_{\mathrm{T}}^{2}=m^{2}
-\frac{\lambda \pi \sin \left(\pi\xi/2\right) }{4\xi L^4m^2R_{\xi}}\int_{U_0 ^{%
\frac{1}{2\xi }}}^{\infty }dx\ \left[ \mathcal{F}_0\left(\xi, U_0,x\right) -\frac{x^2\left( x^{2\xi }-U_0\right) ^{\frac{1}{2}}}{e^{2\pi x} - 1}\right].  \label{mt}
\end{eqnarray}%
The expression above, as expected, goes to zero as $mL\rightarrow\infty$ for any odd value of $\xi$. However, at $mL=0$, that is, in the massless case, it diverges 
for $\xi>1$ as a consequence of infrared divergences that appear due to the Ho{\v{r}}ava-Lifshitz model adopted here. This problem, on the other hand, does not exist 
if we take $\xi=1$, which is the standard case where the Lorentz symmetry is preserved. Hence, for $\xi=1$, Eq. \eqref{mt} can be written as 
\begin{eqnarray}
m_{\mathrm{T}}^{2}=m^{2}+\frac{\lambda m^2}{4\pi^2}\sum_{n=1}^{\infty}\int_{1}^{\infty}dy(y^2-1)^{\frac{1}{2}}e^{-2nmL y},
 \label{mt2}
\end{eqnarray}%
where we have expressed $(e^{2\pi x}-1)^{-1}$ as a sum in $n$ of the exponential above. We have also performed the change of variable $y=xU_0^{-\frac{1}{2}}$, similar to Eq. \eqref{cx1}. Now, by using Eq. \eqref{iden2} we find 
\begin{equation}
m_{\mathrm{T}}^{2}\left( \xi =1\right) =m^{2}\left[ 1+\frac{\lambda }{8\pi
^{2}mL}\sum_{n=1}^{\infty }n^{-1}K_{1}\left( 2nmL\right) \right],
\label{tm1}
\end{equation}%
which is in agreement with known results found in the literature \cite{cruz2020casimir}. Note that the massless scalar field case for the topological mass follows directly from Eq. \eqref{tm1}. 

Nevertheless, in order to better notice the infrared divergences that appear in the massless limit of Eq. \eqref{mt}, let us take the limit $mL\rightarrow 0$ ($U_0\rightarrow 0$) of Eq. \eqref{mt}. For this, we should consider the following expansion for small $mL$: 
\begin{eqnarray}
 \left[(e^{2\pi x }-1) \mathcal{F}_0\left(\xi, U_0,x\right) -x^2\left( x^{2\xi }-U_0\right) ^{\frac{1}{2}}\right]\simeq \frac{\xi x^{2-\xi}}{(\xi-2)}U_0 + O\left(U_0^2\right).
  \label{mt2}
\end{eqnarray}%
Consequently, Eq. \eqref{mt} becomes 
\begin{equation}
m_{\text{\textrm{T}}}^{2}=-\frac{\lambda \pi\sin \left( \pi\xi/2\right)R_{\xi}}{4\left( \xi -2\right) L^2}\int_{0}^{\infty }\frac{dx\ x^{2-\xi }}{\left( e^{2\pi
x}-1\right) }.  \label{topol_mass}
\end{equation}%
As we can see, the integral above does not converge for $\xi>1$. However, for $\xi=1$, we have 
\begin{equation}
m_{\text{\textrm{T}}}^{2}\left( \xi =1\right) =\frac{%
\lambda }{96L^{2}},  \label{r2}
\end{equation}%
which is a known result \cite{cruz2020casimir}. It is clear now that, for any odd values of $\xi$, the expression in Eq. \eqref{mt} is only valid for $m>0$. On the other hand, for $\xi=1$, it provides known results for both the massive and massless scalar field cases.

The generation of topological mass, will reinforce the already strong decay in the vacuum energy density, as mentioned before. In this sense the search of this effect becomes experimentally much harder. 
\begin{figure}[tbp]
\includegraphics[width=.5\textwidth]{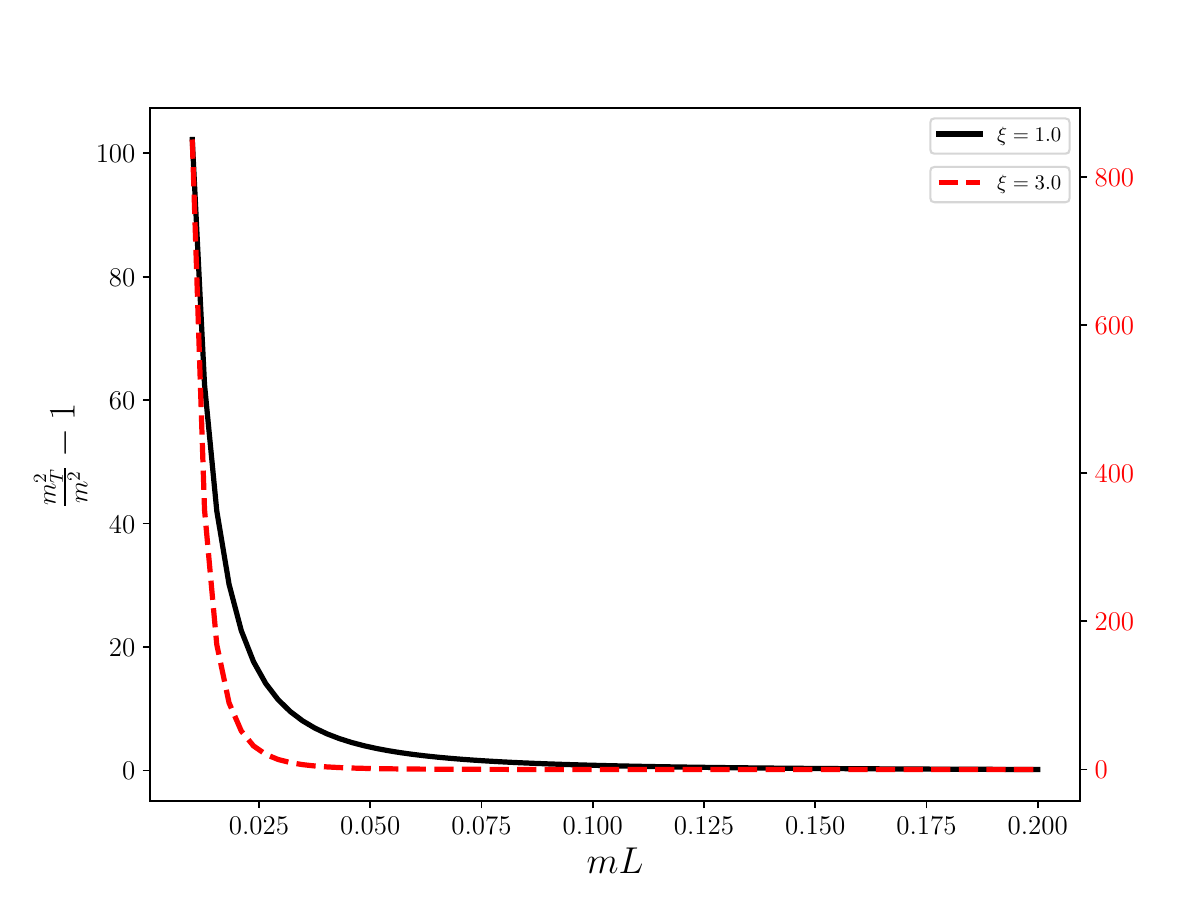}\hfill %
\includegraphics[width=.5\textwidth]{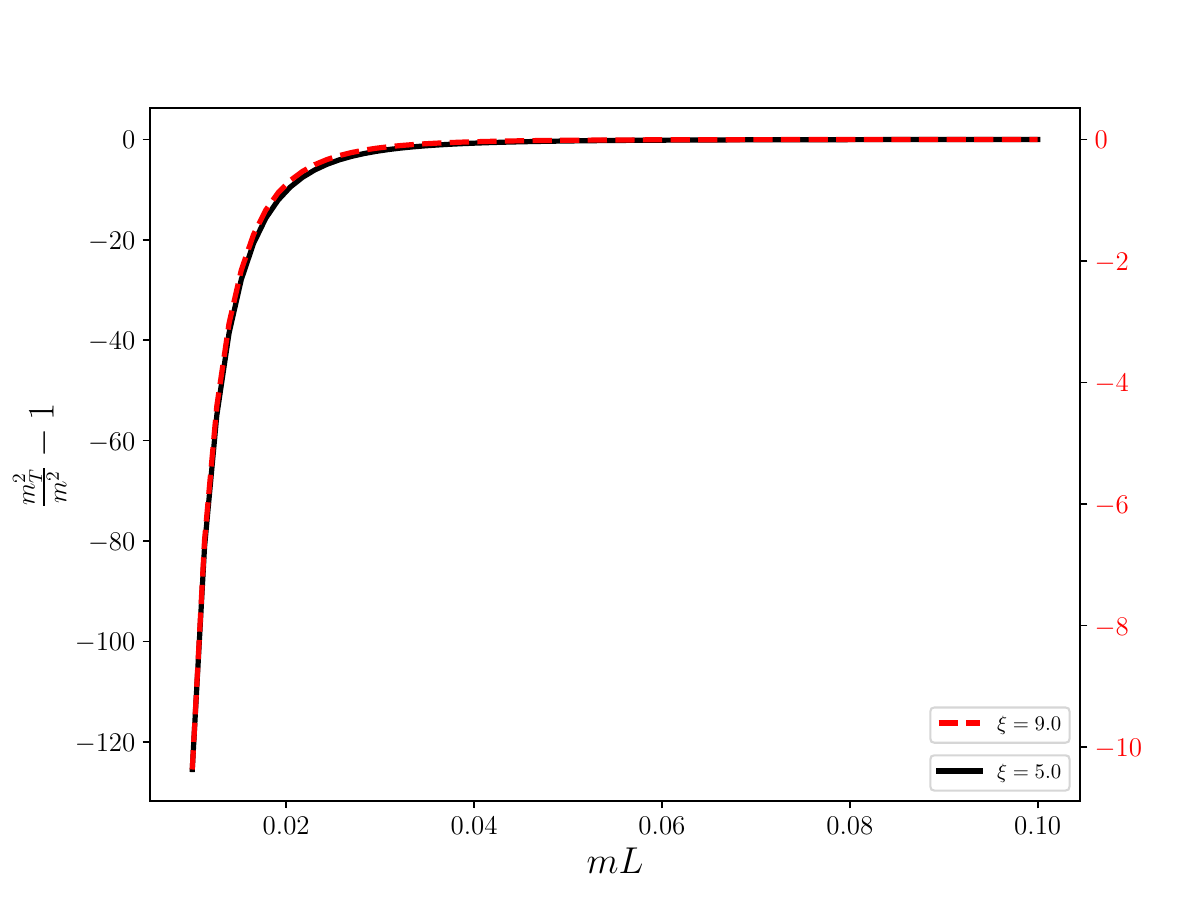}
\caption{Graphs of the dimensionless topological mass $\frac{m_T^2}{m^2}-1$, as a function of $mL$. The graph on
the left considers the values $\protect\xi =1.0$ (solid black curve) and $%
\protect\xi=3.0$ (dashed red curve), while the plot on the right considers
the values $\xi=5.0$ (solid black curve) and $\xi=9.0$ (dashed red curve). The
scale of each curve is shown in the vertical axes with the correspondent color.  
}
\label{fig3}
\end{figure}

In order to plot the expression in Eq. \eqref{mt} with respect to $mL$ it is again useful to use the same change of variable as in Eq. \eqref{variablez}. The correspondent plots are exhibited in Fig.\ref{fig3}. Note that we have plotted only the second term on the r.h.s. of Eq. \eqref{mt}. The graphs clearly show that the topological mass goes to zero as $mL\rightarrow\infty$, as it should be. However, for $mL\rightarrow 0$, the topological mass goes to infinity indicating the presence of infrared divergences in the massless limit.  Moreover, as we can see, for some values of $\xi$ like $\xi=5, 9$, the values of the topological mass become negative which may indicate that an analysis of vacuum stability is necessary within a model where the scalar field considered here interacts with a second field, similar what has been done in Refs. \cite{toms1980interacting, PhysRevD.107.125019}.

\subsection{Estimative for the ratio $\left( l/L\right) ^{\protect\xi -1}$}
\label{sec7}
%
\label{seccomp}In this section we consider the experimental results
presented in \cite{bressi2002measurement} in order to estimate the ratio $%
\left( l/L\right) ^{\xi -1}$. Hence, from Ref.~\cite%
{bressi2002measurement}, we consider%
\begin{equation}
\Delta \nu ^{2}=-\frac{C_{\text{\textrm{Cas}}}}{L^{5}},  \label{cexp}
\end{equation}%
where $C_{\text{\textrm{Cas}}}=\left( 2.34\pm 0.34\right) \times 10^{-28}\ $
Hz$^{2}$m$^{5}$. The connection between the vacuum energy density given in Eq.~(%
\ref{cm01}), with the Casimir energy $E_{\text{\textrm{0}}}$ is given by
\cite{cruz2020casimir},%
\begin{equation}
E_{0}=AL\mathcal{E}_{0},  \label{cpe}
\end{equation}%
where $A$ is the area of the plates. The Casimir force,\ $F_{0}$, is
obtained via derivative of the Casimir energy with respect to the
distance $L$, and the Casimir pressure, $P_{0}$, is the Casimir force
divided by the area $A$, i.e.,%
\begin{equation}
P_{0}=\frac{F_{0}}{A}=-\frac{1}{A}\frac{\partial E_{0}}{\partial L}.
\end{equation}
According to the method proposed in Ref.~\cite{pascoal2008estimate}, we can
write the relation between $\Delta \nu ^{2}$ and $E_{\text{\textrm{%
0}}}$ as,%
\begin{equation}
\Delta \nu ^{2}=-\frac{A}{4\pi ^{2}m_{\text{\textrm{eff}}}}\frac{\partial
P_{0}}{\partial L}=\frac{A}{4\pi ^{2}m_{\text{\textrm{eff}}}}\frac{\partial 
}{\partial L}\left( \frac{1}{A}\frac{\partial E_{0}}{\partial L}\right) ,
\label{vn}
\end{equation}%
where $m_{\text{\textrm{eff}}}$ is an effective mas and the ratio $A/m_{\text{\textrm{eff}}}$ is experimentally estimated in \cite%
{bressi2002measurement}, that is,%
\begin{equation}
\frac{A}{m_{\text{\textrm{eff}}}}\approx 1.746\ \text{Hz}^{2}\text{m}^{3}\text{N}^{-1}.
\label{ra}
\end{equation}
 Note that the expression in Eq.~\eqref{vn} is the general definition of $\Delta \nu ^{2}$ appearing in Eq.~\eqref{cexp}. This is discussed, for instance, in Ref.~\cite{pascoal2008estimate} and will be applied to our case.

Using the result obtained in Eq.~(\ref{cm01}) for the vacuum energy density
in the case of massless scalar field, together with the relation expressed
by Eq.~(\ref{cpe}), we find the vacuum pressure for a massless field in the
Ho{\v{r}}ava-Lifshitz model considered here as%
\begin{equation}
P_{0}=-\sin \left( \frac{\pi \xi }{2}\right) \frac{\left( 2+\xi \right)
l^{\left( \xi -1\right) }}{2^{\xi +4}\pi ^{2}}\Gamma \left( \xi +2\right)
\zeta _{R}\left( \xi +3\right) \frac{1}{L^{3+\xi }}.
\end{equation}%
Therefore the quantity $\Delta \nu ^{2}$ defined in Eq.~(\ref{vn}) can be
written in the form,%
\begin{equation}
\Delta \nu ^{2}=-\frac{A}{m_{\text{\textrm{eff}}}}\sin \left( \frac{\pi \xi 
}{2}\right) \frac{\left( 3+\xi \right) \left( 2+\xi \right) }{2^{\xi +6}\pi
^{4}}\Gamma \left( \xi +2\right) \zeta _{R}\left( \xi +3\right) \left( \frac{%
l}{L}\right) ^{\xi -1}\frac{1}{L^{5}}.
\end{equation}%

Considering the estimated result written in Eq.~(\ref{ra}), together with
the Eq.~(\ref{cexp}), we write the ratio $\left( l/L\right) ^{\xi -1}$ in
the following way%
\begin{equation}
\left( \frac{l}{L}\right) ^{\xi -1}=0.34\times 10^{-28}\frac{m_{\text{%
\textrm{eff}}}}{A}\frac{2^{\xi +6}\pi ^{4}}{\left( 3+\xi \right) \left(
2+\xi \right) }\frac{1}{\sin \left( \frac{\pi \xi }{2}\right) \Gamma \left(
\xi +2\right) \zeta _{R}\left( \xi +3\right) },  \label{lxi}
\end{equation}
where we have considered the error bar of $C_{\text{\textrm{Cas}}}$. Hence, 
from the experimental value of $A/m_{\text{\textrm{eff}}}$ written in Eq.~(%
\ref{ra}), the expression in Eq. (\ref{lxi}) allows us to construct Table \ref{t1}. 
\begin{table}[tbp]
\begin{tabular}{|l|l|l|l|l|}
\hline\hline
$\xi$ & $\xi=3$ & $\xi=5$ & $\xi=7$ & $\xi=9$ \\ \hline
$\left( \frac{l}{L}\right) ^{\xi-1} $ & $\left( \frac{l}{L}\right) ^{2}
=1.32588\times 10^{-27}$ & $\left( \frac{l}{L}\right) ^{4} =9.59569\times
10^{-29}$ & $\left( \frac{l}{L}\right) ^{6} =4.27789\times 10^{-30} $ & $%
\left( \frac{l}{L}\right) ^{8} =1.2973\times 10^{-31}$ \\ \hline\hline
\end{tabular}%
\caption{ $\left( l/L\right) ^{\protect\xi -1}$ as a function of $\protect%
\xi $.}
\label{t1}
\end{table}
Note that we are taking the absolute value of the ratio $l/L$, since it is a
positive quantity. As we can see from Table \ref{t1}, the bigger the value of $\xi$
the smaller the ratio $l/L$. The biggest value is for $\xi=3$, which is about $l/L\simeq 3.6\times 10^{-14}$.

We emphasize that the analysis in Ref.~\cite{bressi2002measurement} makes no assumptions about the nature of the quantum field. The data are simply better fitted by the electromagnetic quantum modes responsible for the Casimir effect. However, the error bar in $C_{\text{Cas}}$ below Eq. \eqref{cexp} allows for the inclusion of other contributions, which could improve constraints on the Casimir force, including potential effects from additional quantum field modes.

Furthermore, we point out that the experimental analysis in Ref.~\cite{bressi2002measurement} was not designed to test Lorentz symmetry violations, being instead based on the standard $L^{-5}$ dependence for parallel plates. However, as previously noted, the experimental uncertainty in $C_{\text{Cas}}$ may accommodate potential Lorentz-violating effects. We may therefore obtain a conservative estimate for $l/L$ by assuming such effects lie within the experimental error margin. A dedicated experimental study would be required to establish more stringent constraints.

\section{Concluding remarks}

\label{sec5}
 The main objective of this paper was to analyze the vacuum energy and the generation of topological mass for a system composed by a massive self-interaction scalar quantum field in the scenario of a Ho{\v{r}}ava-Lifshitz-like model, and in this way to extend previous analysis developed in \cite{maluf2020casimir}. In this context, the vacuum energy was calculated considering that the field obeys  Dirichlet boundary condition on two large parallel plates separated by a distance $L$. The energy spectrum is derived and using the $\zeta-$ function regularization approach we have calculated the renromalized effective potential, $V_{\text{eff}}^{\text{R}}(\Phi)$, given in \eqref{vr} for an arbitrary value of the critical exponent $\xi$. The vacuum energy density is obtained from this potential in the vacuum state. Our result is presented in \eqref{cas}.  We have explicitly shown that for even values of this parameter, this one-loop correction vanishes. Unfortunately, an analytical expression is not possible for general values of $\xi $. Moreover, the two-loop correction to the vacuum energy  was presented in \eqref{c1} for arbitrary value $\xi$, in the massive field case. Its massless limit, on the other hand,
has been obtained in Eq.~(\ref{c2}) and only gives a finite result for $\xi=1$, the standard case
where the Lorentz symmetry is preserved. However, for $\xi>1$ the integral does not converge as a consequence of infrared divergences introduced by the model.

 Another relevant calculation was developed in \ref{sectop}, that is, the generation of topological mass. This observable is given by the second derivative with respect to the field $\Phi$ of the renormalized effective potential, prepared in the vacuum state. The corresponding expression was given in \eqref{mt} as an integral representation for arbitrary value of $\xi$. We explicitly show that it goes to zero in the limit $mL\to\infty$ and vanishes for even value of the critical exponent.

In Fig.\ref{fig1} we have exhibited the graphs of the dimensionless vacuum energy density $E\left( mL\right) =4\pi
^{2}L^{4}\mathcal{E}_{0}$, as a function of $mL$,
considering the values $\xi =1$, $3$, $5$, $7$. Additionaly, the vacuum energy density for the case of a
massless field is presented in Eq.~(\ref{cm01}), which shows the dependecy
on $L$ and $\xi $, and reproduces the obtained results in Ref.~\cite%
{ulion2015casimir}. 

In Fig.\ref{fig2} we have exhibited graphs for the dimensionless two-loop contribution to the
vacuum energy density $\Delta E\left( mL\right) =\frac{128}{\lambda \pi ^{2}}%
L^{4}\mathcal{E}_{0}^{\lambda }$ as a function of $mL$, considering the
values $\xi =1,3,5,7$. The graphs show that the correction goes to zero
as $mL\rightarrow\infty$ (as it should be) and diverges as $mL\rightarrow 0$ (for $\xi>1$),
indicating the infrared divergences.

In the case of a massive field, its mass acquires a correction which
depends on the critical exponent, $\xi $, on the parameter $L$ and also on
the self-interaction coupling constant $\lambda $. This correction is
presented in Eq.~(\ref{mt}). However, for $\xi>1$, this expression for the topological
mass diverges as $mL\rightarrow 0$, also indicating the presence of infrared divergences. 
This can be better seen in Eq. (\ref{topol_mass}), which clearly converges only for $\xi=1$. In contrast,
for $mL\rightarrow\infty$ the topological mass goes to zero, as expected. This is shown in the graphs of Fig.\ref{fig3}.
The graph on the right shows that for $\xi=5,9$ the topological mass is negative and a vacuum stability analysis
must take place in a model with interacting fields.

We have also provided an estimative for the ratio $l/L$ by comparison with
experimental results presented in Ref.~\cite{bressi2002measurement}. The
values of the ratio $l/L$ are presented in Table \ref{t1}. It shows that the ratio 
$l/L$ decreases as $\xi$ increases, with the biggest value being about $l/L\simeq 3.6\times 10^{-14}$. As a future work,
we plan to look further into the issue of infrared divergences that appear in the massless scalar field case, and also
analyze a system where two scalar fields interact. 

It is worth emphasizing that the scalar field model considered in this work serves purely as a toy model for probing general features of Lorentz-violating effects on confined quantum fields. While scalar theories offer a technically tractable setting, it is well known that the extension of Lorentz-violating structures to electromagnetic fields is highly nontrivial due to the stringent constraints imposed by gauge invariance (see Ref.~\cite{Kostelecky:2009zp}). The admissible Lorentz-breaking operators in gauge theories differ significantly from those in scalar models, both in structure and physical implications. Studies involving HL-like theories with gauge and spinor fields confirm that each sector requires a specific treatment to preserve renormalizability and/or gauge invariance, as discussed for instance in Refs.~\cite{Farias:2011jc, Farias:2012ed}. Therefore, the results presented here should be viewed as qualitative indications rather than direct predictions for electromagnetic phenomena.
 
Finally, in contrast to previous studies that focused on free scalar fields, in this work we incorporate a self-interaction potential and compute loop corrections to the vacuum energy up to two-loop order. We also investigate the generation of topological mass and provide phenomenological estimates based on experimental data. These aspects constitute significant extensions to prior HL-like scalar models, and uncover new features such as infrared divergences for \(\xi > 1\) and dependence of the Casimir-like pressure on the interaction strength. Although our model is scalar and serves as a simplified toy model, it offers valuable qualitative insights into how Lorentz-violating structures may influence quantum vacuum effects.

{\acknowledgments} The author H.M. is partially supported by the
Brazilian agency National Council for Scientific and Technological
Development (CNPq) under Grant No. 308049/2023-3.

\bigskip

\bigskip

\appendix

\section{Abel-Plana formula}

\renewcommand{\theequation}{A.\arabic{equation}} \setcounter{equation}{0}%
\label{app1} In this appendix we use the Abel-Plana formula \cite%
{bordag2009advances,PhysRevD.21.933,
ulion2015casimir,saharian2007generalized},%
\begin{equation}
\sum_{n=1}^{\infty }f\left( n\right) =\int_{0}^{\infty }dx\ f\left( x\right)
-\frac{1}{2}f\left( 0\right) +i\int_{0}^{\infty }dx\frac{f\left( ix\right)
-f\left( -ix\right) }{e^{2\pi x}-1}\ ,
\label{ABF}
\end{equation}%
to perform the sum in $n$ present in Eq.~(\ref{zh1}). Hence, the generalized zeta
function in Eq.~(\ref{zh1}) can be written as a sum of three contributions, namely,%
\begin{equation}
\zeta \left( s\right) =\zeta _{\mathrm{M}}\left( s\right) +\zeta _{\mathrm{1P%
}}\left( s\right) +\zeta _{0}\left( s\right).  \label{z}
\end{equation}%
The first contribution, $\zeta _{\mathrm{M}}\left( s\right) $, reads, 
\begin{eqnarray}
&&\zeta _{\mathrm{M}}\left( s\right) =\frac{\Omega _{3}\pi ^{\frac{1}{2}+\xi
-2\xi s}}{8\xi b^{s-\frac{1}{2}}L^{2+\xi -2\xi s}}\frac{\Gamma \left( s-%
\frac{1}{2}-\frac{1}{\xi }\right) \Gamma \left( s-\frac{1}{2}\right) }{%
\Gamma \left( s\right) \Gamma \left( s+\frac{1}{2}-\frac{1}{\xi }\right) } 
\notag \\
&&\times \int_{0}^{\infty }dx\left( x^{2\xi }+U\right) ^{\frac{1}{2}+\frac{1%
}{\xi }-s}\,_{2}F_{1}\left( 1-\frac{1}{\xi },s-\frac{1}{2}-\frac{1}{\xi },s+%
\frac{1}{2}-\frac{1}{\xi },\frac{U}{x^{2\xi }+U}\right) \ ,  \label{abel1}
\end{eqnarray}%
which is in fact the Minkowski contribution that comes from 
the integral in the first term on r.h.s. of Eq. \eqref{ABF}. The second 
contribution comes from the second term on r.h.s. of Eq. \eqref{ABF}
and is associated with the one plate case. This
contribution is given by
\begin{eqnarray}
&&\zeta _{\mathrm{1P}}\left( s\right) =-\frac{\Omega _{3}\pi ^{\frac{1}{2}%
+\xi -2\xi s}U^{\frac{1}{2}+\frac{1}{\xi }-s}}{16\xi b^{s-\frac{1}{2}%
}L^{2+\xi -2\xi s}}\frac{\Gamma \left( s-\frac{1}{2}\right) }{\Gamma \left(
s\right) }  \notag \\
&&\times \frac{\Gamma \left( s-\frac{1}{2}-\frac{1}{\xi }\right) }{\Gamma
\left( s+\frac{1}{2}-\frac{1}{\xi }\right) }\,_{2}F_{1}\left( 1-\frac{1}{\xi 
},s-\frac{1}{2}-\frac{1}{\xi },s+\frac{1}{2}-\frac{1}{\xi },1\right) \ .
\label{abel2}
\end{eqnarray}%

For the third contribution we have to separate the integral in the variable $x$ into
two intervals: from $[0,\ U^{\frac{1}{2\xi }}]$ and from $[U^{\frac{1}{2\xi }%
},\ \infty )$. For the first interval and considering $\xi $ an integer
number, there is no contribution. In the second interval and considering $\xi 
$ an integer number, the situation is more delicate. In this case\ we have $%
U^{\frac{1}{2\xi }}<x$, which yields,%
\begin{eqnarray}
&&f\left( ix\right) -f\left( -ix\right) =2i\sin \left( \frac{\pi \xi }{2}%
+\pi -\pi \xi s\right) \left[ x^{2\xi }+\left( -1\right) ^{\xi }U\right] ^{%
\frac{1}{2}+\frac{1}{\xi }-s}  \notag \\
&&\times \,_{2}F_{1}\left( 1-\frac{1}{\xi },s-\frac{1}{2}-\frac{1}{\xi },s+%
\frac{1}{2}-\frac{1}{\xi },\frac{U}{U+\left( -1\right) ^{\xi }\left(
x\right) ^{2\xi }}\right) .
\end{eqnarray}%
From the above expression we can infer that for $\xi $ even, the difference
above goes to zero, since we will eventually have to take $s\rightarrow 0$.
Additionally, since the third term in the Abel-Plana formula is the one
which originates the vacuum energy density, we see that for $\xi $ even we
have a vanishing vacuum energy density. Hence, we will investigate only the cases in
which $\xi $ is an odd number. Thus, for the third contribution in Eq. \eqref{z}
that comes from the integral in the third term on the r.h.s. of Eq. \eqref{ABF}, we have
\begin{eqnarray}
&&\zeta _{0}\left( s\right) =-\frac{\Omega _{3}\sin \left[ \pi \left( \frac{%
\xi }{2}+1-\xi s\right) \right] }{4\xi \pi ^{2\xi s-\frac{1}{2}-\xi }b^{s-%
\frac{1}{2}}L^{2+\xi -2\xi s}}\frac{\Gamma \left( s-\frac{1}{2}-\frac{1}{\xi 
}\right) \Gamma \left( s-\frac{1}{2}\right) }{\Gamma \left( s\right) \Gamma
\left( s+\frac{1}{2}-\frac{1}{\xi }\right) }  \notag \\
&&\times \int_{U^{\frac{1}{2\xi }}}^{\infty }dx\frac{\left( x^{2\xi
}-U\right) ^{\frac{1}{2}+\frac{1}{\xi }-s}}{e^{2\pi x}-1}\,_{2}F_{1}\left( 1-%
\frac{1}{\xi },s-\frac{1}{2}-\frac{1}{\xi },s+\frac{1}{2}-\frac{1}{\xi },%
\frac{U}{U-x^{2\xi }}\right) \ .  \label{abel33}
\end{eqnarray}%
Thus, we have the generalized zeta function, Eq.~(\ref{z}), as a sum of three
terms, namely, Eq.~(\ref{abel1}), (\ref{abel2}) and (\ref{abel33}).

The next step is to evaluate the generalized zeta function and its
derivatives at $s=0$, for each of the three terms separately. All
the three terms are zero at $s=0$, and by series expansion about $s=0$ we
obtain the derivatives as%
\begin{eqnarray}
&&\zeta _{\mathrm{M}}^{\prime }\left( 0\right) =\frac{\Omega _{3}\pi ^{\xi
+1}b^{\frac{1}{2}}U^{\frac{1}{2}+\frac{3}{2\xi }}}{4\left( \xi +2\right)
L^{\xi +2}}\int_{0}^{\infty }dy\left( y^{2\xi }+1\right) ^{\frac{1}{2}+\frac{%
1}{\xi }}F\left( 1-\frac{1}{\xi },-\frac{1}{2}-\frac{1}{\xi },\frac{1}{2}-%
\frac{1}{\xi },\frac{1}{1+y^{2\xi }}\right) ,  \notag \\
&&\zeta _{\mathrm{1P}}^{\prime }\left( 0\right) =-\frac{\Omega _{3}\pi ^{\xi
+\frac{1}{2}}b^{\frac{1}{2}}U^{\frac{1}{2}+\frac{1}{\xi }}}{16\xi b^{\frac{1%
}{\xi }}L^{\xi +2}}\Gamma \left( -\frac{1}{2}-\frac{1}{\xi }\right) \Gamma
\left( \frac{1}{\xi }\right) ,  \notag \\
&&\zeta _{0}^{\prime }\left( 0\right) =\sin \left( \frac{\pi \xi }{2}\right) 
\frac{\Omega _{3}\pi ^{1+\xi }b^{\frac{1}{2}}}{\left( \xi +2\right) L^{2+\xi
}}\sum_{n=1}^{\infty }\int_{U^{\frac{1}{2\xi }}}^{\infty }dx\mathcal{F}%
\left( n,\xi ,U,x\right) .
\end{eqnarray}%
Note that in the first line of the above equation, we change the integration
variable as $y^{2\xi }=x^{2\xi }/U$. Besides, this term is divergent and
represents the contribution with no boundaries (no parallel plates, that is,
the Minkowski contribution) and it should be discarded \cite%
{ulion2015casimir}. The contribution from the second line of the above
equation is the contribution coming from one plate, which grows with
positive powers of the mass present in $U$ (see Eq. \eqref{u01}) and, thus should be discarded.
Finally, the third line gives the contribution which truly 
generates the vacuum energy density.


\end{document}